# Phase Change Logic via Thermal Cross-Talk for Computation in Memory

Nadim Kanan, Raihan Sayeed Khan, Zachary Woods, Helena Silva, Ali Gokirmak

*Abstract*— We have computationally demonstrated logic function implementations using multi-contact phase change devices integrated with CMOS circuity. Read terminals are isolated from each other utilizing amorphized regions formed between different pairs of write contacts, allowing implementation such as multiplexers and 2x2 routers. Thermal cross-talk during the write operations is utilized to recrystallize the previously amorphized regions to achieve toggle operations. Integration of these multi-contact phase change elements with CMOS circuity reduce the CMOS area needed by ~50% for logic functions such as toggle-multiplexing or JK-multiplexing, with the added benefit of non-volatility. An electro-thermal modeling framework with dynamic materials models are used to capture the device dynamics, and current and voltage requirements.

## I. Introduction

The electronics market is continuing to grow, as functionality increases for reduced cost, and the market expands into the developing world with mobile applications [1]–[3]. Annual shipment of new cellular phones is exceeding 1.8 billion [4] (1.4 billion are smart phones) [5], [6] and tablet PCs surpassed 200 million [7] as print media is being replaced by electronic alternatives [8]–[11]. The demand for overall computer and communication infrastructure is also increasing as a result.

Today, typical data-heavy computations are limited by memory access latencies due to the I/O bottleneck to DRAM (dynamic random access memory), flash memory and hard drive [12] [13]. If large amounts of high-speed non-volatile memory (100s of GBs) could be integrated atop the CPU, the need for off-chip DRAM, hard drive (storage) and even motherboard could be eliminated for some applications. Monolithic integration of memory and logic can be realized as computation-in-memory of larger-scale applications. These computer-on-chip and computation-in-memory approaches can achieve significant speed improvements at a fraction of the power for data intensive computations.

Silicon CMOS is still at the forefront of scaling and conventional logic, driving process technologies [14], which also enables integration of complementary nanoscale resistive non-volatile memory (RRAM) memory technologies such as phase change memory (PCM) [15],[16] (Fig. 1).

RRAM structures are compact 2-terminal devices that utilize reversible changes in resistance of a small volume of material through magnetic switching (magneto-resistance) as in magnetic RAM (MRAM) [17], electro-thermal effects (PCM, memristors) or electro-chemical effects (memristors). The three common approaches to achieve large resistance contrast for non-magnetic RRAM devices are [18]:

*1)* changing a materials phase between amorphous and crystalline (PCM) [13], [19]–[23],

*2)* formation of metallic filaments in non-conductive thin layers [24]–[27] (memristors) or

*3)* generation and annihilation of oxygen vacancies in metal oxides [28]–[33] (memristors).

All RRAM devices experience joule heating [34]–[39], electrical breakdown [40], thermal transport [41], [42], thermoelectric effects [34], [35], [43], [44], electro-migration [45], [46] and percolation transport [32], [33], [47], which make them harder to model compared to conventional electronic devices [48]–[52].

Typical PCM devices are vertical mushroom cells (Fig. 1) composed of a bottom-contact, a shared top-contact, and a phase change material that offers a large resistivity ($\rho$) contrast between the amorphous (a-GST) and crystalline (x-GST) phases, such as $Ge_2Sb_2Te_5$ (GST) [44], [53]–[56]. The set and reset operations are achieved through localized self-heating via short voltage pulses ($V_{pulse}$) to transition between the set (crystalline) and reset (amorphous) states. These devices can

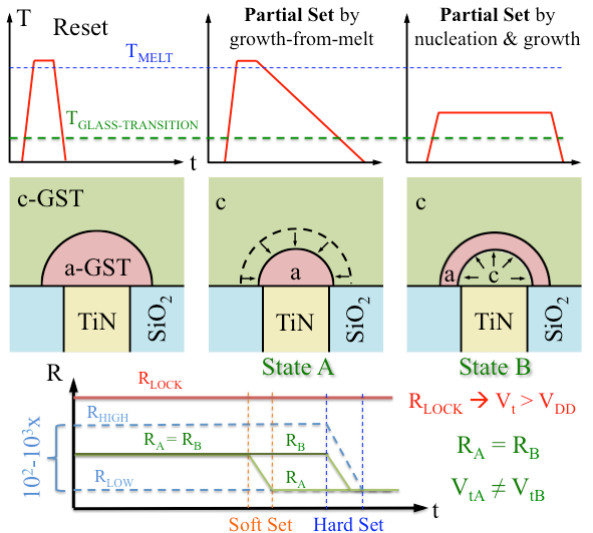

**Fig. 1.** PCM mushroom cells are reset (to high resistance) using short duration high amplitude pulses (left). Set operation (going back to low-resistance) can be achieved by re-melting and growing-from-melt (center) or by annealing to nucleate and grow (right). An intermediate resistance value can be achieved in two ways ($R_A = R_B$) with sufficiently different threshold voltages ($V_t$) for the next set operation. A soft-set can flip state A, and a hard-set can flip all cells other than the over-reset (locked) cells ($V_t > V_{DD}$).

*Authors performed the presented research activities at University of Connecticut, Storrs, CT USA, with the support of US National Science Foundation grant # ECCS 1150960 and ECCS 1711626.

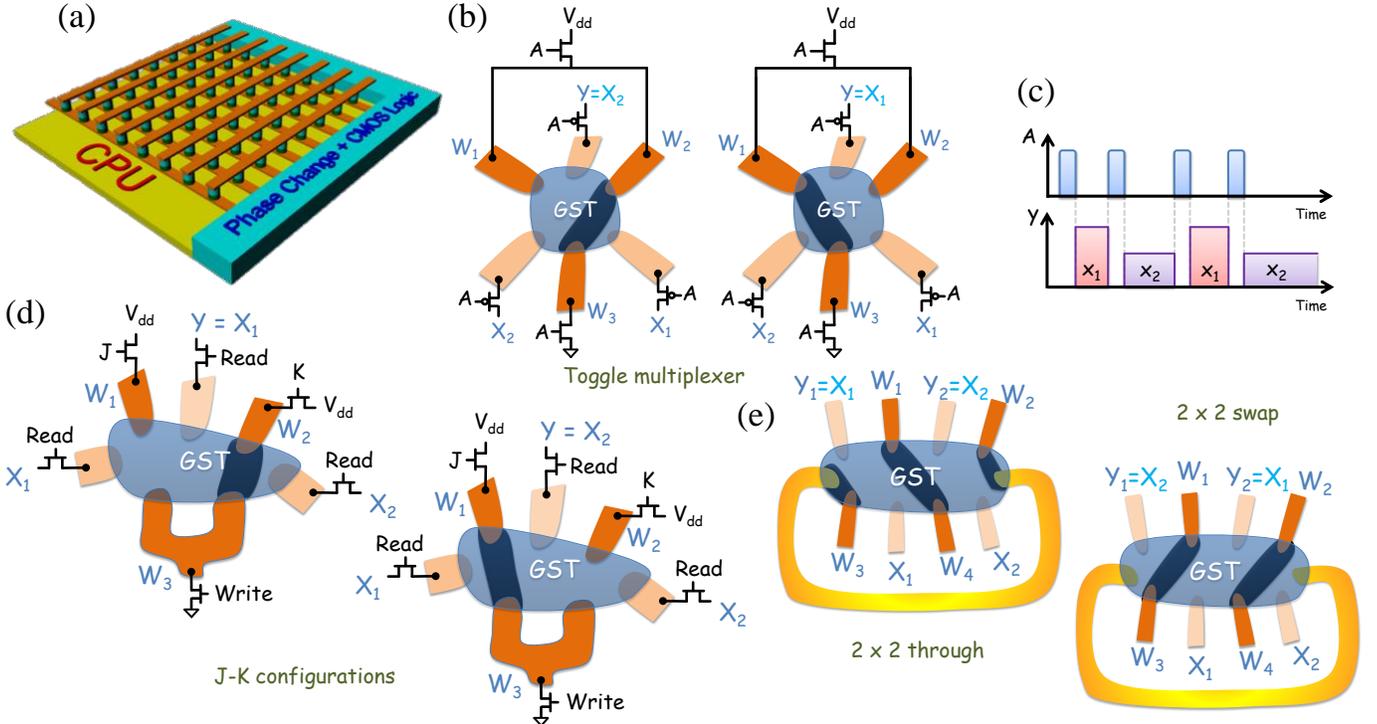

Fig. 2. (a) Schematic showing phase change memory integrated on top of CPU with phase change and CMOS combined logic at the interface. (b)-(e) Schematic top-view of lateral multi-contact phase change logic elements. Melting and amorphization between the write terminals (W) are used to isolate sections of the phase change patch ($Ge_2Sb_2Te_5$, GST) to route the signals. (b) shows the toggle (T-) multiplexer/flip-flop configurations: Activating A would cause one of the write paths ($W_1W_3$ or $W_2W_3$) to melt isolating a read path and routing the other signal. A subsequent pulse in A would melt the other write path while the rest of the patch crystallizes due to thermal cross-talk. (b) offers deterministic initial state as $W_2$ is closer to $W_3$ compared to $W_1$. (c) shows the input pulse and the output at y terminal. The functionality can be further extended to J-K filp-flop (d): activation of $W_2$ & $W_3$ isolates $X_2$ after the write cycle, leaving Y connected to $X_1$ (Y = $X_1$). Similarly Y = $X_2$ can be achieved by activating $W_1$ & $W_3$. (e) shows 2 x 2 routers that are configured by sequential activation of W pairs. Any input-output combination, including 'swap' ($Y_1 = X_2$, $Y_2 = X_1$) can be achieved if $S_1$ & $S_2$ are connected.

provide $\sim 10^2$-$10^4$x contrast in total device resistance [21] (k$\Omega$ to M$\Omega$ range), cycled $> 10^{12}$ times [57] (flash memory can be cycled up to $\sim 10^6$ times) and retain data for $> 10$ years at CPU temperature. Memory window increases over time due upward drift of the $\rho_{a\text{-GST}}$ [58]–[60]. This resistance drift limits the implementation of multi-bit-per-cell implementation of conventional PCM cells. Cells with shunt resistances are demonstrated to overcome this difficulty [57]. PCM devices can be integrated with CMOS at the back-end-of-the-line as many layers atop CMOS circuitry using low temperature processes as done in 3D XPoint memory [61].

## II. MULTI-CONTACT PHASE CHANGE LOGIC DEVICES

Emergence of PCM as a viable memory technology has sparked interest in its implementation as switches [62] as well as components of neural networks complementing CMOS circuitry [21]. Complementing CMOS circuitry with functional phase change elements in the memory layer to achieve memory control, multiplexing, routing and neuromorphic computations will relieve the area concerns in the underlying CMOS layer, making it easier to realize computer-on-chip and computation-in-memory as well as hardware implementation of artificial neural networks (Fig. 2).

The multi-contact phase change devices described in this manuscript can be used to implement flip-flops [63], routers[64], [65], multiplexers [66], counters and state machines at a smaller CMOS footprint as discussed below.

These non-conventional phase-change devices can be integrated with CMOS access devices to offer more functionality and/or reduced area compared to conventional nonvolatile memory devices and CMOS circuitry, and can also enable computations using intermittent power.

Multi-contact phase change elements offer the possibility to implement functions and reconfigurable routing, going beyond conventional PCM devices. The main drawback for *all phase change logic,* similar to other non-CMOS approaches, is the inability to amplify the signal back to the supply level ($\pm V_{dd}$). However, complementing CMOS with multi-contact phase-change elements offers significant advantages in logic circuit footprint (area) and standby power. The multi-contact phase change elements described here utilize

*1)* localized heating to amorphize portions of a phase change patch to isolate read contacts,

*2)* thermal cross-talk to crystallize an amorphous region when a nearby area is being amorphized [67]–[70].

A family of lateral (Fig. 2) and vertical phase change logic devices and circuits can be implemented using these processes. The basic circuit configuration and operation of a toggle (T-) multiplexer is depicted in Fig. 2b,c and Fig. 3. This 6-contact device, interfaced with 5 MOSFETs, isolates the output (Y) from one of the inputs ($X_1$, $X_2$) by amorphizing a region between two write contacts each time a write signal (A) is received (Fig. 2c). The geometry of each amorphized region is affected by the thermal losses and the current path, which tends to form a filament due to thermal runaway as polycrystalline phase change materials become more conductive at elevated temperatures [71]–[73]. When the cell is written the very first time, two parallel high current-density paths are formed between top and bottom write terminals. If the two top-contacts are shorted as shown (Fig. 2b), the path which heats up slightly more experiences thermal runaway and melts (Fig. 3b). Upon rapid termination of the write pulse, the molten path becomes amorphous, isolating one of the read contacts (Fig. 3c). When the same write pulse is applied again, current predominantly flows through the second path, which was not amorphized previously (Fig. 3d). The write pulse is kept long enough to recrystallize the previously amorphized region. Hence, when the second write pulse is terminated, the second current path is amorphized and the first path is crystallized (Fig. 3e). The interaction between the two paths through thermal cross-talk gives rise to the toggle operation. The device remains in the 'read' state whenever it is not receiving a write pulse, controlled by the CMOS access circuitry. We demonstrate the toggle operation of the 6-contact device in a circular geometry using our electro-thermal finite-element modeling framework with dynamic materials [50]–[52], [74]. The access devices, electrical waveforms, device geometry and thermal losses have to be designed appropriately to achieve the desired operation.

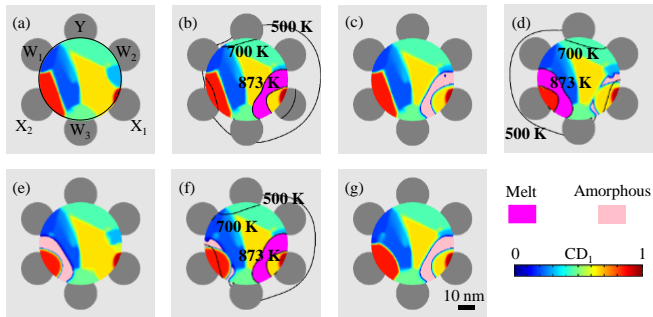
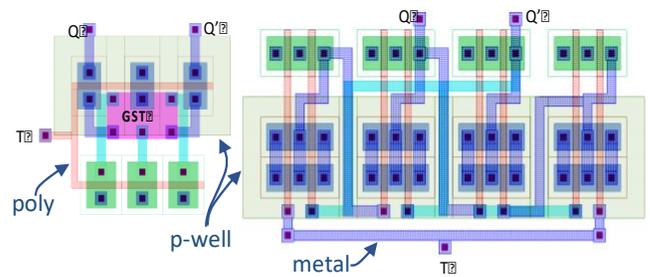

**Fig. 3.** Crystallinity profiles and temperature contours as the device is cycled with $W_1$, $W_2$, and $W_3$. Grain orientation is plotted in light rainbow, amorphous in peach, and melt in pink. The same signal is applied to $W_1$ and $W_2$ for each melt/quench cycle. Initially the heating is symmetric, but a single path soon dominates. After the device cools, the next pulse is applied and current flows through the more conductive crystalline path, while thermal crosstalk between the two paths crystallizes the previously amorphous area.

**Fig. 4.** Layouts illustrating the 50% area reduction using the phase-change T-flip-flop using 6 transistors (left) compared to the conventional T-flip-flop built with 4 NAND gates (right). Both offer the same logic functionality.

It is possible to achieve a deterministic initial state by breaking the symmetry either through increased series resistance for one of the write contacts or an asymmetric phase change device geometry. One example of such an asymmetric device is shown in (Fig. 2d. Here, two separate paths of different lengths are defined between two pairs of write contacts. These 7-contact structures can also be used to realize a JK flip-flop if the two of the write terminals are accessed by independently controlled MOSFETs (Fig. 2d).

Implementation of the 6-contact phase-change logic + CMOS as a toggle multiplexer yields ~50% less footprint compared to its conventional CMOS counterpart (Fig. 4), accounting for the increased widths for the MOSFETs used for write access. The conventional CMOS JK-multiplexer requires 18 transistors while the PCM-JK multiplexer requires only 6 transistors, corresponding to approximately 66% savings in footprint. The PCM-logic alternatives come with the added advantage of non-volatility, which makes these circuits good alternatives for intermittent power applications (sensor networks, etc.) where conventional CMOS circuits would need frequent write / read access to non-volatile memory to successfully complete the computation between unpredictable intermittent power periods.

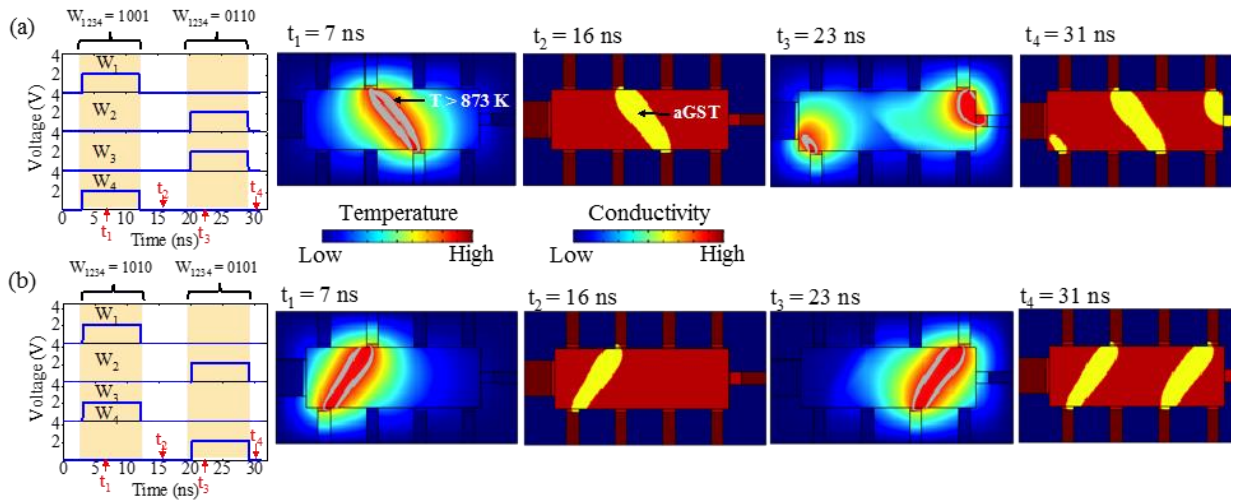

**Fig. 5.** Sequence of write pulses applied and the corresponding thermal profiles and resistivity maps at the indicated times. Simulated thermal profile during a write pulse (left) and the resulting resistivity map during the read cycle (right) for a planar 10-contact router. All contacts other than the left contact is designed to be 10 nm. The side contacts are electrically short circuited through a wire. Sizing one of the side contacts wider than the other is desired to decrease the electrical resistance. Blocking of the narrower side contact with amorphized GST is sufficient to electrically isolate the two sides. Electro-thermal model with effective media approximation used for these simulations. In the first set of write pulses (a) electrical paths in GST between $W_1$ & $W_4$ and $W_2$ & $W_3$ are amorphized one at a time with high speed sequential pulses. Device is set to $Y_1=X_1$, $Y_2=X_2$ state. In the second set of electrical pulses (b) electrical paths between $W_1$ & $W_3$ and $W_2$ & $W_4$ are amorphized setting the device to 'swap' $Y_1=X_2$, $Y_2=X_1$ state.

Each lateral phase change element can also be configured to have a larger number of contacts as in the case of the 10-contact structure shown in Fig. 2e. Connecting the two side contacts ($S_1$ & $S_2$) together yields a 2 x 2 router that is configured by 4 contacts (Fig. 5), an alternative to a cylindrical structure [75]. This device can be configured with high-speed sequential pulses to amorphize the paths between the top and bottom contact pairs, avoiding recrystallization of the paths formed earlier. Reconfiguration can be achieved by keeping the first pulse sufficiently long to recrystallize the areas amorphized in the last pulse sequence. The snapshots in Fig. 5 show the thermal profiles during configuration and the resulting resistivity maps corresponding to the applied pulse sequences. Here, the two configurations implement $\{Y_1=X_1, Y_2=X_2\}$ (through) and $\{Y_1=X_2, Y_2=X_1\}$ (swap). The wider contact used on the left-side to reduce electrical resistance.

A subsection of these 10-contact structures (Fig. 3c) can be used to implement toggle operations as in Fig. 3a,b with nearest neighbor interactions. Larger structures with phase change strips with large number of contacts on two sides can also be envisioned for this purpose.

While established functions can be implemented using these device concepts, they are intended to be enabling technologies rather than become a replacement technology, and hence go beyond what is typically achievable by integration of non-volatile memory elements with CMOS.

It is possible to achieve faster operation than what is demonstrated in the above examples by using AIST (higher growth velocity) [76] instead of GST, thermal engineering of the structure, sizing the access devices and engineering the write pulses. While it is typically assumed that reset operation requires melting, it is possible to amorphize grain boundaries and interfaces at temperatures substantially below bulk melting temperature ($T_{MELT}$) [77]. It is also possible to amorphize at lower temperatures by substantial hole injection at steep thermal gradients or junctions, through impact ionization by high voltage short duration pulses [53].

III. RELIABILITY AND ENDURANCE

PCM cells have been demonstrated with endurance up to >$10^{12}$ cycles [57] while $10^{15}$-$10^{17}$ cycles of endurance is necessary to achieve logic operations at the CPU clock speed. The typical PCM cell failures are due to void formation and elemental segregation. The lateral devices described here differ from the compact two-terminal vertical cells in two ways that make them more resilient: (i) devices utilize thermal cross-talk, hence dielectric breakdown is not required for recrystallization and the structure does not experience fields as strong as conventional cells that drive elemental segregation, (ii) these lateral structures deliver the same functionality even if there is a high-density of voids in the phase change material (Fig. 7). The devices can also be perforated by etching a pattern formed by block-copolymers [78].

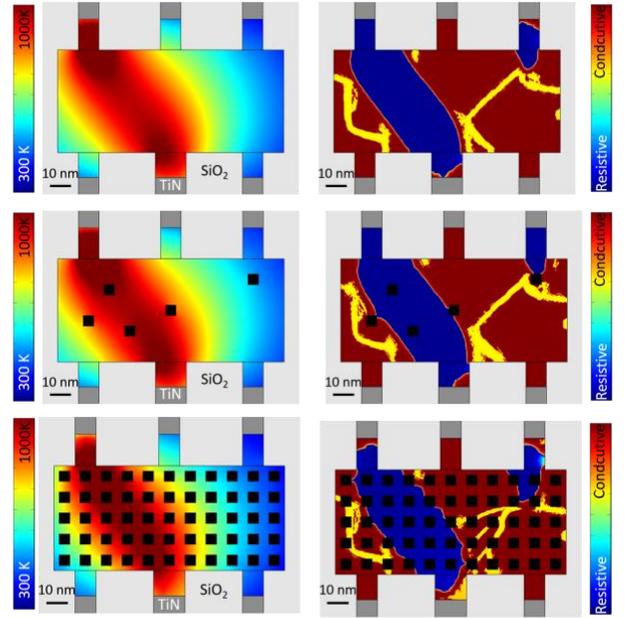

**Fig. 6.** Thermal profile at peak temperature (left) and conductivity maps after cool-down for a defect-free GST patch (top), with voids as shown in black (center) and perforated GST film. All devices achieve the same functionality. Perforation reduces the switching power by ~30%. Grain boundary physics is included in the electro-thermal model. Grain boundaries are clearly visible in conductivity map.

IV. MODELING FRAMEWORK

The electro-thermal model we have constructed to simulate phase-change device operation self-consistently solves the current continuity (1) and heat transfer (2) equations, using *COMSOL Multiphysics* finite-element tool [71]:

$$\nabla \cdot J = -\nabla \cdot \underbrace{\sigma(T)\nabla V}_{Ohm's\ Law} - \nabla \cdot \underbrace{\sigma(T)\ S\nabla T}_{Seebeck\ current} = 0 \quad (1)$$

$$\underbrace{d_G C_P(T)\frac{dT}{dt}}_{Heating} - \underbrace{\nabla \cdot (\kappa(T)\nabla T)}_{Heat\ diffusion} = \underbrace{\frac{J^2}{\sigma(T)}}_{Joule\ heating} - \underbrace{J\ T\ \nabla S}_{Thomson\ heat} \quad (2)$$

where J is the current density, V is the electric potential, S is the Seebeck coefficient, and $d_G$ is the mass density. This model accounts for the thermoelectric effects through the introduction of the Seebeck current term in (1) and the Thomson heat term in (2) [79]. We include thermal boundary resistances to account for reduced thermal conductivity at the interfaces and a field-dependent electrical conductivity component for phase-change material to capture dielectric breakdown (set operation) [49]. The electro-thermal model and the materials' parameters are described in detail in our recent publications [52], [71], [77], [80]. Access transistors are integrated with the finite element simulations by implementing the basic nFET circuit model available in COMSOL [81].

The dynamic materials modeling is handled through a rate equation (3) using published nucleation rates and growth velocities, either using a course mesh modeling a large volume, or using ~ 1nm mesh to capture discrete nucleation:

$$\frac{dCD}{dt} = Nucleation\,(T, CD, random) + Growth\,(T, CD) - Amorphization(T, CD) \quad (3)$$

This approach is significantly more efficient compared to discrete nucleation models using individual domains to define the material characteristic [82], [83].

We have demonstrated that the thermoelectric contribution is substantial, hence there is a significant impact of the device polarity through our computational studies, as was also experimentally demonstrated by others [84]. Changing the contact material does not only impact cell resistance and thermal losses[85], but also the thermoelectric heating of the active area[49], [86]–[89]. The out of plane thickness used for simulations is 20 nm.

## V. SUMMARY

Thermal cross-talk in phase change devices can be utilized as a coupling mechanism to achieve logic functions and decrease the CMOS footprint necessary for high density PCM arrays. Multi-contact devices utilizing thermal-cross talk integrated with CMOS allow rail-to-rail operation and significantly reduce the area necessary to implement logic functions and routing for memory access. The switching speed of these devices are comparable to PCM switching speeds and these devices are suitable for memory control and routing. Hence, the presented approach to reduce CMOS resources can enable computer-on-chip implementations with very high-density non-volatile storage monolithically integrated with logic in a single chip.